\documentclass[conference]{IEEEtran}
\IEEEoverridecommandlockouts

\usepackage{cite}
\usepackage{amsmath,amssymb,amsfonts}
\usepackage{graphicx}
\usepackage{textcomp}
\usepackage{xcolor}
\usepackage{multirow}
\usepackage{url}
\usepackage{booktabs}

\IEEEoverridecommandlockouts        
\def\BibTeX{{\rm B\kern-.05em{\sc i\kern-.025em b}\kern-.08em
    T\kern-.1667em\lower.7ex\hbox{E}\kern-.125emX}}

\begin{document}

\title{\LARGE ProDCARL: Reinforcement Learning-Aligned Diffusion Models for De Novo Antimicrobial Peptide Design}


\author{\IEEEauthorblockN{Fang Sheng}
\IEEEauthorblockA{
\textit{University of Toronto}\\
Toronto, Canada \\
fang.sheng@mail.utoronto.ca}
\and
\IEEEauthorblockN{Mohammad Noaeen}
\IEEEauthorblockA{
\textit{University of Toronto}\\
Toronto, Canada \\
m.noaeen@utoronto.ca}
\and
\IEEEauthorblockN{Zahra Shakeri}
\IEEEauthorblockA{
\textit{University of Toronto}\\
Toronto, Canada \\
zahra.shakeri@utoronto.ca}
}

\maketitle

\begin{abstract}
Antimicrobial resistance threatens healthcare sustainability and motivates low-cost computational discovery of antimicrobial peptides (AMPs).
De novo peptide generation must optimize antimicrobial activity and safety through low predicted toxicity, but likelihood-trained generators do not enforce these goals explicitly.
We introduce ProDCARL, a reinforcement-learning alignment framework that couples a diffusion-based protein generator (EvoDiff OA-DM 38M) with sequence property predictors for AMP activity and peptide toxicity.
We fine-tune the diffusion prior on AMP sequences to obtain a domain-aware generator.
Top-k policy-gradient updates use classifier-derived rewards plus entropy regularization and early stopping to preserve diversity and reduce reward hacking.
In silico experiments show ProDCARL increases the mean predicted AMP score from 0.081 after fine-tuning to 0.178.
The joint high-quality hit rate reaches 6.3\% with pAMP $>$0.7 and pTox $<$0.3.
ProDCARL maintains high diversity, with $1-$mean pairwise identity equal to 0.929.
Qualitative analyses with AlphaFold3 and ProtBERT embeddings suggest candidates show plausible AMP-like structural and semantic characteristics.
ProDCARL serves as a candidate generator that narrows experimental search space, and experimental validation remains future work.

\end{abstract}


\begin{figure*}[!t]
    \centering
    \includegraphics[width=\textwidth]{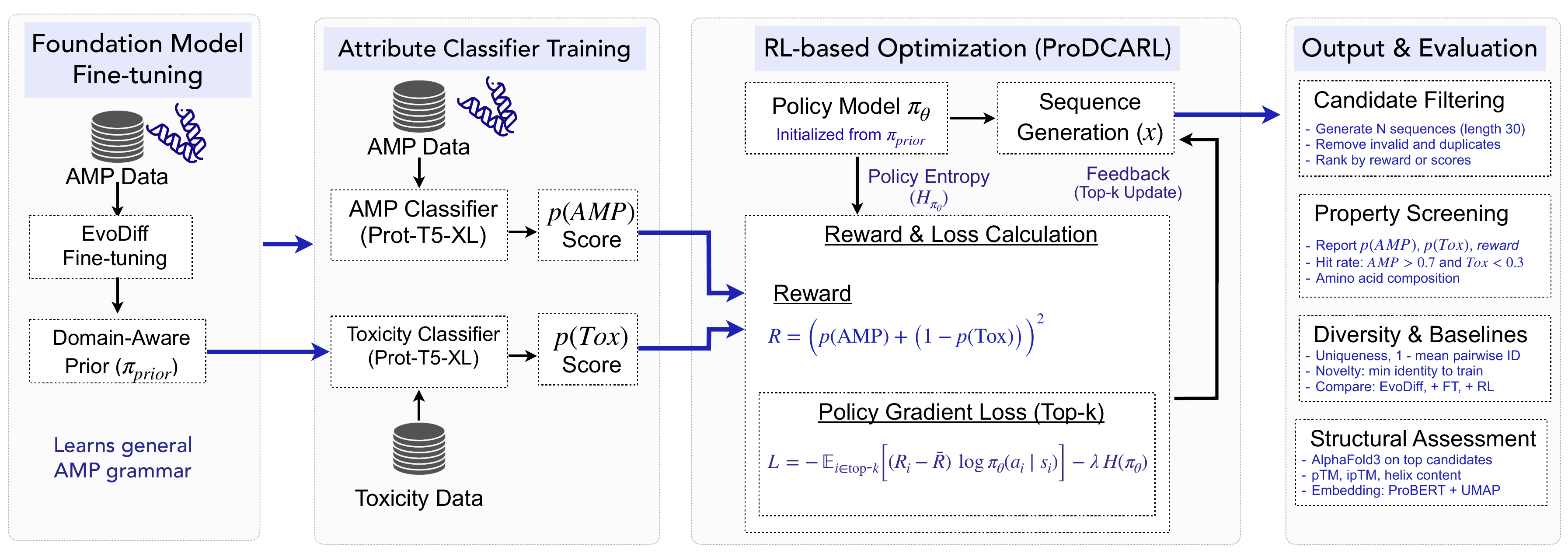}
    \caption{Overview of the ProDCARL pipeline. The framework integrates data preparation, classifier training, EvoDiff prior fine-tuning, and reinforcement learning alignment with rewards.}
    \label{fig:pipeline}
    \vspace{-5mm}
\end{figure*}

\section{Introduction}
Antimicrobial resistance is a growing threat to global health and has created an urgent need for new therapeutic strategies beyond traditional antibiotics. Antimicrobial peptides (AMPs), typically 10–50 amino acids in length, have become promising alternatives to conventional antibiotics \cite{hancock2006antimicrobial, wang2019antimicrobial}. These short peptides target bacterial membranes or intracellular components \cite{hancock2006antimicrobial, wang2019antimicrobial} and exhibit broad-spectrum activity with a reduced tendency to induce resistance \cite{mookherjee2020antimicrobial, mahlapuu2016antimicrobial}. However, the discovery of novel AMPs through traditional experimental pipelines is costly and time-consuming \cite{koczulla2003antimicrobial}. As a result, researchers have increasingly turned to computational methods, particularly deep learning techniques, to accelerate AMP discovery \cite{porto2018silico, santos2024discovery}.

Recent advances in protein Large-Language Models (PLMs) have enabled effective learning of protein sequence representations \cite{heinzinger2024sty, rives2021biological, lin2022language}. Generative modeling approaches such as variational autoencoders, generative adversarial networks, and autoregressive models have shown promise in protein design \cite{madani2020progen, alley2019unified, repecka2021expanding}. Diffusion probabilistic models \cite{sohl2015deep, ho2020denoising} have become a powerful paradigm in protein sequence generation, exemplified by the EvoDiff system \cite{alamdari2023protein}. However, most existing generative methods optimize sequence likelihood or structural plausibility alone, rather than directly optimizing functional objectives such as high antimicrobial activity and low toxicity. This misalignment can yield candidates that appear structurally plausible but prove ineffective or unsafe as therapeutics.

Reinforcement learning (RL) offers a strategy to align generative models with specific biochemical objectives \cite{popova2018deep, olivecrona2017molecular}. In drug discovery, deep RL techniques have guided de novo molecule generation toward desired properties by using predictive models as reward functions \cite{popova2018deep, olivecrona2017molecular}. A recent study applied RL to explore peptide sequence space and generated drug-like peptide candidates \cite{wang2024reinforcement}. This outcome demonstrates the potential of RL-driven optimization in peptide design. Nevertheless, RL-based generation can suffer from mode collapse and reward exploitation if not properly constrained. In such cases, the model may converge to a narrow set of sequences or find unrealistic shortcuts to maximize the reward. These challenges highlight the need for approaches that maintain diversity while steering generation toward therapeutically relevant outcomes.

In this work, we propose ProDCARL, a framework that integrates a diffusion-based protein generator with reinforcement learning to directly optimize AMP activity and safety. We fine-tune a pretrained diffusion model (EvoDiff \cite{alamdari2023protein}) on a collection of AMP sequences to create a domain-specific generator. We then couple this generator with two discriminative reward models that evaluate antimicrobial activity and toxicity, respectively. These reward models leverage a ProtT5-XL protein language model \cite{elnaggar2021prottrans} and incorporate additional convolutional and self-attention layers \cite{vaswani2017attention}. Each classifier outputs a predicted AMP probability (pAMP) and a predicted toxicity (pTox) for each sequence. We use these values as feedback signals to guide the generator. We train the generative model with a top-$k$ policy gradient algorithm that emphasizes high-scoring samples. To preserve sequence diversity and prevent reward hacking, we include an entropy regularization term and an early stopping criterion in the training loop. To our knowledge, ProDCARL is the first approach to combine diffusion-based sequence generation with reinforcement learning for antimicrobial peptide design.

Our experiments show that ProDCARL outperforms diffusion-only baselines in antimicrobial peptide design. After reinforcement learning alignment, the mean predicted AMP activity score of generated peptides more than doubled (from 0.081 after diffusion fine-tuning to 0.178). Moreover, 6.3\% of generated peptides qualify as high-quality hits. We define a high-quality hit as a sequence that simultaneously achieves high predicted activity (pAMP $>$ 0.7) and low predicted toxicity (pTox $<$ 0.3). This hit rate represents a substantial improvement in the yield of desirable candidates. Importantly, these gains come with only a minimal loss of diversity. The generated peptide set retains high diversity, with $1-\text{mean pairwise identity}=0.929$, which indicates that the optimization does not collapse to near-duplicates under the reported early-stopped model. Qualitative analyses further support the relevance of the generated peptide candidates. Structure predictions using AlphaFold and sequence embeddings from ProtBERT suggest that the top-ranked candidates from ProDCARL adopt plausible AMP-like folds and sequence patterns. ProDCARL represents a significant advance in computational antimicrobial peptide design by bridging diffusion-based generative modeling with reinforcement learning to accelerate the discovery of new peptide therapeutics. 

The implementation of ProDCARL is available at:
\url{https://github.com/HIVE-UofT/ProDCARL}.

\section{Methods}

\subsection{Problem Formulation}
Figure~\ref{fig:pipeline} represents the overall pipeline of ProDCARL. In this pipeline, we generate peptide sequences $x \in A^{L}$, where $A$ is the amino-acid alphabet and $L = 30$.
Let $p_{\mathrm{AMP}}(x)$ and $p_{\mathrm{Tox}}(x)$ denote outputs from the AMP-activity and toxicity predictors, respectively.
We use these outputs as ranking scores for screening rather than calibrated probabilities.
We optimize the following scalar reward to favor high activity with low predicted toxicity:
\begin{equation}
R(x) = \left( p_{\mathrm{AMP}}(x) + \left(1 - p_{\mathrm{Tox}}(x)\right) \right)^{2},
\end{equation}
We update the parameters $\theta$ of the generative policy $\pi_{\theta}$ to increase $\mathbb{E}_{x \sim \pi_{\theta}}[R(x)]$.

We optimize the top-$k$ policy-gradient objective with an entropy term during training as follows:
\begin{equation}
\mathcal{L}
= - \mathbb{E}_{i \in \mathrm{top}\text{-}k}\!\left[\left(R_i - \bar{R}\right)\,\log \pi_{\theta}(x_i)\right]
- \lambda\, H(\pi_{\theta}),
\end{equation}
Here $R_i$ denotes the reward for sequence $i$, and $\bar{R}$ denotes the batch mean reward baseline.
The term $H(\pi_{\theta})$ is the mean token predictive entropy across positions, and $\lambda$ is the entropy weight, and we set $\lambda=3$ in our experiments.
We update $\theta$ with AdamW at a learning rate of $2\times 10^{-4}$ for all runs.
Top-$k$ updates use the top $30\%$ of sequences to concentrate gradients on informative and high-reward samples.
Early stopping triggers when monitoring metrics detect diversity collapse during training in the sampling distribution.

\begin{table}[t]
\caption{Dataset summary after residue filtering, length filtering, and redundancy reduction.}
\label{tab:dataset}
\centering
\begin{tabular}{lcc}
\hline
Task and label & Train & Test \\
\hline
AMP positive (APD3+DBAASP) & 95{,}209 & 48{,}777 \\
AMP negative (SwissProt reviewed) & 100{,}018 & 51{,}517 \\
Toxic positive (CSM-Toxin+ToxinPred2+ATSE) & 2{,}117 & 374 \\
Toxic negative (UniProt reviewed, filtered) & 5{,}995 & 1{,}058 \\
\hline
\end{tabular}
\vspace{-3mm}
\end{table}

\subsection{Data Preparation}
Two datasets were used in this study: AMP, and toxic peptide sequences. AMP sequences were collected from APD3\cite{wang2016apd3} and DBAASP \cite{pirtskhalava2021dbaasp}. Negative samples were randomly drawn from SwissProt \cite{uniprot2019uniprot} by selecting reviewed sequences without known antimicrobial annotations. We also removed any sequence that contained non-canonical residues to standardize the input alphabet. 
To reduce homologous leakage across splits, we clustered sequences with CD-HIT at 70\% identity.
We then split the data at the cluster level, so each cluster stayed in one split.
We randomly assigned 30\% of clusters to the final test set.
After clustering and splitting, the training set contained 95{,}209 positives and 100{,}018 negatives.
The test set contained 48{,}777 positives and 51{,}517 negatives after the same procedure (Table~\ref{tab:dataset}).

For toxicity prediction, we adopt the dataset construction described in CAPTP \cite{jiao2024integrated}, which integrates toxic peptides from CSM-Toxin \cite{morozov2023csm}, ToxinPred2 \cite{sharma2022toxinpred2}, and ATSE \cite{wei2021atse}. We sample non-toxic sequences from UniProt reviewed entries \cite{uniprot2019uniprot} using the query \texttt{NOT KW-0800 AND NOT KW-0020 AND reviewed:true}. We reduce redundancy with CD-HIT at $90\%$ identity \cite{fu2012cd} and remove sequences longer than 50 residues. The final set contains 2{,}491 toxic peptides and 7{,}053 non-toxic peptides, and we reserve 15\% of samples for testing.

\subsection{Classifier Models}
Two classifiers were implemented, one for AMP activity and one for toxicity. Both models shared the same architecture and were built on a pretrained Prot-T5-XL encoder \cite{elnaggar2021prottrans} to retrieve global sequence representations. Hidden representations were passed to multi-scale convolutional filters (kernel sizes of 3, 5, and 7 respectively; 100 channels each). The resulting representation was fused by self-attention layers (encoded dimension: 300; num\_head:4; num\_layer:2) and finally passed to multi-layer perceptrons (dimension of 150, 2 respectively; dropout: 0.3) for classification.

As shown in Table~\ref{tab:test_performance}, both classifiers achieved strong discriminative performance on held-out test sets, which provide reliable reward signals to guide the reinforcement learning process. Because AMP and toxicity datasets are class imbalanced and may include label noise, we report ROC-AUC and PR-AUC. During RL, these predictors serve as surrogate reward models. To limit over-optimization to a single predictor, we use entropy regularization, top-k updates, deduplication filtering, and early stopping when diversity collapse is observed.

We train both classifiers with a binary negative log-likelihood objective using a \texttt{sigmoid} output layer. We optimize with \texttt{AdamW} and use early stopping based on validation PR-AUC computed on a held-out subset of the training split. We report ROC-AUC and PR-AUC on the final test set because both tasks are class imbalanced. Each predictor output, $p_{\mathrm{AMP}}(x)$ and $p_{\mathrm{Tox}}(x)$, is used as a screening score and not as a calibrated probability. We use batch size $B_{\mathrm{cls}}=\,$200, learning rate $1\times10^{-4}$, weight decay $1\times10^{-5}$, and patience $3$ epochs.

\begin{table*}[htbp]
\centering
\caption{Test performance of AMP and Toxicity models}
\label{tab:test_performance}
\resizebox{.6\textwidth}{!}{
\begin{tabular}{lccccc}
\toprule
Model & Accuracy & Sensitivity & Specificity & ROC-AUC & PR-AUC \\
\midrule
AMP Model      & 0.973 & 0.907 & 0.974 & 0.979 & 0.851 \\
Toxicity Model & 0.888 & 0.837 & 0.908 & 0.950 & 0.899 \\
\bottomrule
\end{tabular}%
}
\end{table*}

\begin{figure}[h] 
    \centering
    \includegraphics[width=\columnwidth]{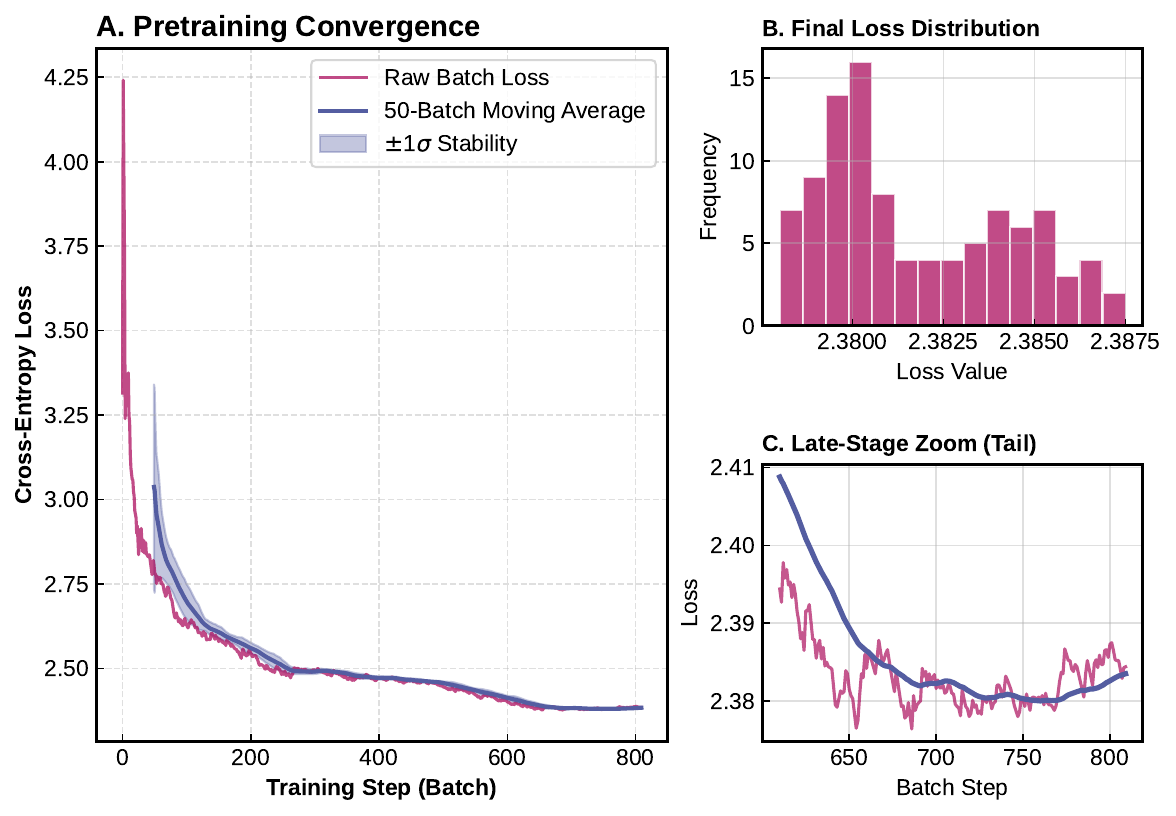}
    \vspace{-3mm}
    \caption{\textbf{Quantitative analysis of diffusion prior fine-tuning.} (a) Training trajectory for negative log-likelihood loss and a 50-batch moving average with $\pm 1\sigma$ shading. (b) Distribution of loss values for the final 100 batches. (c) Late-stage loss values for steps 650-800.}
    \label{fig:pretrain}
\end{figure}

\subsection{Diffusion Prior}
We use EvoDiff \cite{alamdari2023protein} as the generative backbone and select the order-agnostic denoising model (OA-DM 38M). The forward process applies random masking noise, and the denoiser predicts masked residues from the unmasked context. We fine-tune OA-DM on AMP positives from the AMP training split to obtain a domain-aware prior $\pi_{\mathrm{prior}}$. We optimize a negative log-likelihood objective over masked positions under the EvoDiff masking schedule.

To summarize training behavior in Figure~\ref{fig:pretrain}, the pretraining loss decreased steadily from about 4.25 to a plateau near 2.38. The variance envelope and late-stage zoom show stable performance in the final 200 batches, with no pronounced stochastic oscillations typical of large-scale protein language model fine-tuning. Since cross-entropy can be interpreted as the negative log-likelihood of predicting the correct residue, the reduction corresponds to improvement in residue prediction accuracy. The model has only $\sim3\%$ token-hit rate at initialization and exceeding $\sim10\%$ accuracy after convergence. 

To match the training trajectory in Figure~\ref{fig:pretrain}, we fine-tuned for 800 gradient steps and fixed the sequence length to $L=30$ for fine-tuning and sampling. We report the token hit rate as the fraction of masked positions where the \texttt{argmax} residue matches the ground-truth residue.

\subsection{Reinforcement Learning Loop}
After fine-tuning, a reinforcement learning (RL) loop was employed to further align peptide production with low toxicity and high AMP activity. During the $50$ RL iterations, the pretrained EvoDiff model samples a batch of $200$ candidate peptide sequences of length $30$.  These sequences were then assigned scores using AMP and toxicity classifiers. A scalar reward was computed as
\begin{equation}
R = \left(p_{\text{AMP}} + (1 - p_{\text{Tox}})\right)^2
\end{equation}
which combined AMP activity and toxicity scores through squared (quadratic) scaling to amplify the reward signal.  

For diffusion-based generation, we use the OA-DM denoiser, which outputs a categorical distribution over amino acids at each position in the final denoising step.
We treat this distribution as the policy $\pi_\theta$ and approximate the sequence log-probability by summing token log-probabilities across positions:
\begin{equation}
\log \pi_\theta(x) \approx \sum_{j=1}^{L} \log \pi_\theta(x_j \mid s_j),
\end{equation}
Here, $s_j$ denotes the model conditioning at position $j$ in the final denoising step.
This approximation does not compute the full diffusion-trajectory likelihood, but it provides a stable and differentiable surrogate for policy-gradient fine-tuning toward downstream objectives.

We compute the entropy term as
\begin{equation}
H(\pi_\theta)=\frac{1}{BL}\sum_{b=1}^{B}\sum_{j=1}^{L}\left[-\sum_{a\in\mathcal{A}}\pi_\theta(a\mid s_{b,j})\log\pi_\theta(a\mid s_{b,j})\right],
\end{equation}
where $B$ denotes the batch size of sampled sequences and $\mathcal{A}$ denotes the amino-acid alphabet. We trigger early stopping when the diversity metric decreases for $T_{\mathrm{pat}}$ consecutive RL iterations. We set $T_{\mathrm{pat}}=\{3\}$ in all reported runs.

To avoid overfitting, we run RL for up to 50 iterations and report the early-stopped checkpoint at iteration 40.


\subsection{Reproducibility and Responsible Use}
All models are evaluated \textit{in silico} on held-out test splits defined at the CD-HIT cluster level to reduce homologous leakage.
The predictor outputs are surrogate screening scores and do not replace experimental assays. To narrow the experimental search space and reduce resource-intensive screening, this work prioritizes computational triage over direct biological claims.
Any clinical or therapeutic conclusion requires wet-lab validation and an appropriate biosafety review. Training and evaluation scripts, along with the processed splits, will be released upon publication. All code, processed splits, and training scripts are available at \url{https://github.com/HIVE-UofT/ProDCARL}.

\section{Results and Discussion}

\subsection{Evaluation Protocol and Metrics}
For each method (i.e., EvoDiff, EvoDiff+FT, EvoDiff+RL, ProDCARL), the corresponding generator produced fixed-length peptide sequences ($L = 30$). The analysis excluded sequences with non-canonical residues and removed exact duplicates before summary statistics. The reported metrics include (i) Avg AMP $= \mathrm{mean}\, p_{\mathrm{AMP}}(x)$, (ii) Avg Tox $= \mathrm{mean}\, p_{\mathrm{Tox}}(x)$, (iii) hit rate $=$ the fraction meeting joint thresholds ($p_{\mathrm{AMP}} > 0.7$ and $p_{\mathrm{Tox}} < 0.3$), and (iv) diversity $= 1 -$ mean pairwise sequence identity over the evaluated set. Since predictor outputs serve as surrogate screening scores rather than calibrated probabilities, these metrics support relative comparisons under a consistent evaluation protocol.

\begin{figure} 
    \centering
    \includegraphics[width=\columnwidth]{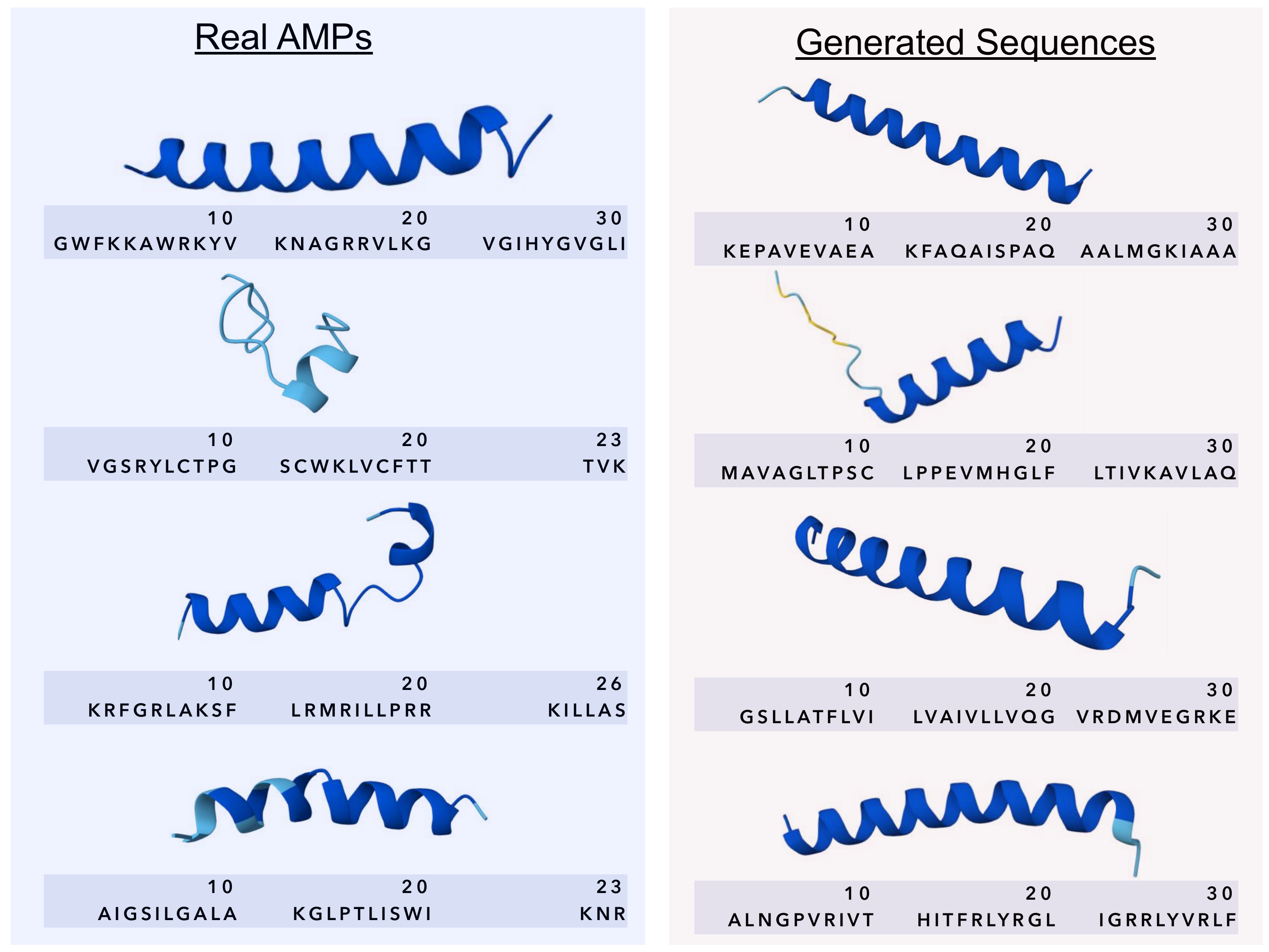}
    \vspace{-3mm}
    \caption{\textbf{Qualitative structural comparison of natural and ProDCARL-generated peptides (AlphaFold3).} Several generated candidates show predominantly $\alpha$-helical conformations consistent with common AMP motifs. Predictions are used as qualitative plausibility checks for short peptides.}
    \label{fig:cand}
\end{figure}
\subsection{Qualitative Structural Plausibility (AlphaFold3)}
To obtain qualitative structural hypotheses for the top 10 generated sequences (ranked by pAMP), we predicted 3D conformations using AlphaFold3 \cite{abramson2024accurate}. Four candidates exhibited predominantly  $\alpha$-helical conformations (Figure~\ref{fig:cand}), a motif frequently associated with many membrane-active AMPs. The remaining candidates showed mixed coil/helix structure, which may reflect either (i) peptides that are partially disordered in solution and adopt structure upon membrane interaction, or (ii) sequences that satisfy the activity predictor but are less structurally constrained. Because many short peptides are conditionally structured, we treat AlphaFold3 outputs as plausibility checks rather than functional validation \cite{hancock2006antimicrobial}.


\subsection{Representation-Space Analysis (ProtBERT + UMAP)}
\begin{figure}[h] 
    \centering
    \includegraphics[width=\columnwidth]{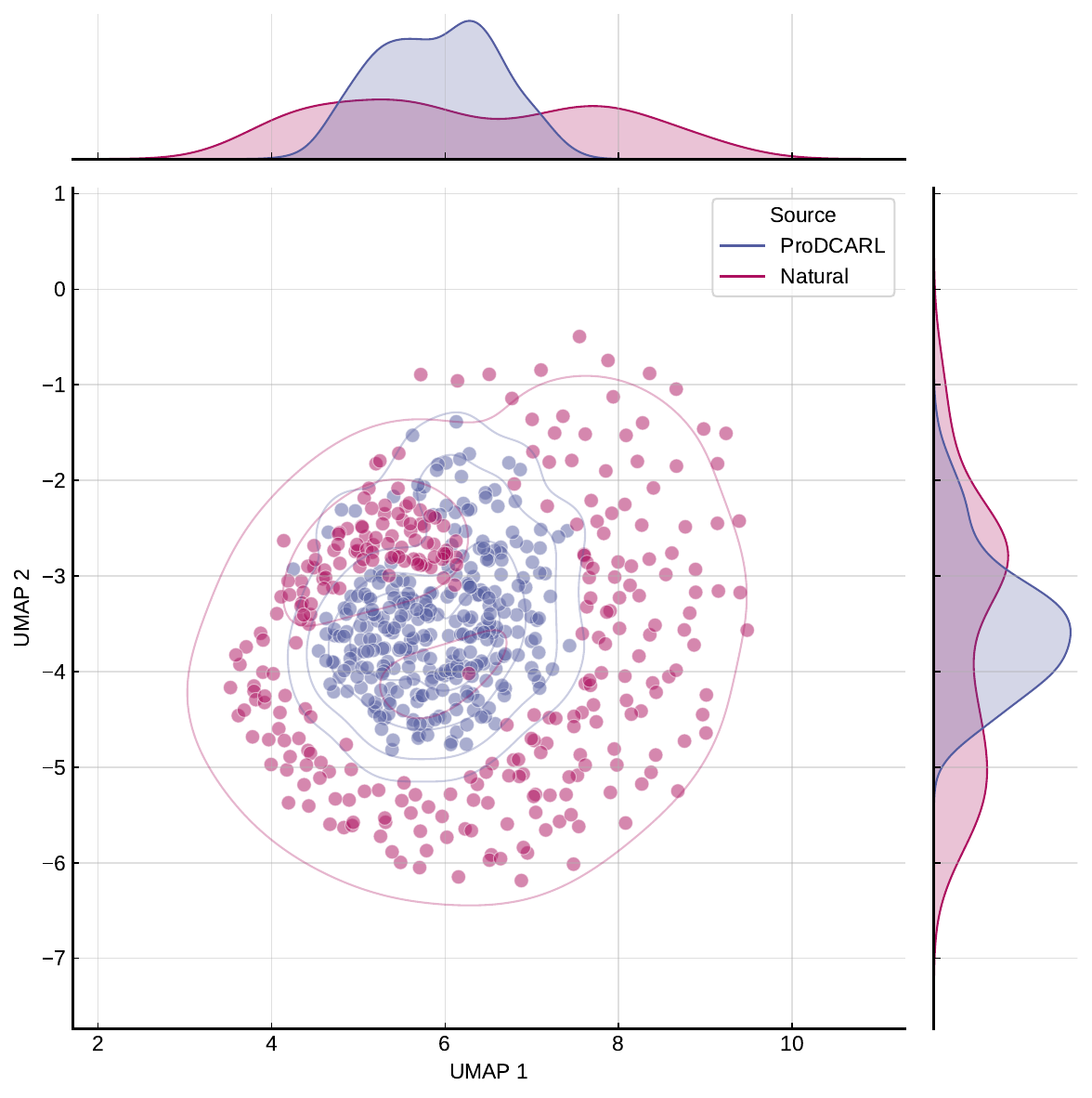}
    \vspace{-2mm}
    \caption{\textbf{ProtBERT-UMAP visualization comparing ProDCARL-generated peptides to natural AMPs (APD).} The overlap suggests alignment in representation space, while the narrower spread of generated candidates likely reflects fixed length and reward shaping rather than the full natural AMP diversity.
}
    \label{fig:umap}
\end{figure}

To test whether generated candidates occupy an AMP-like region of protein representation space, we embedded sequences with ProtBERT \cite{elnaggar2021prottrans} and visualized them with UMAP \cite{mcinnes2018umap} (Figure~\ref{fig:umap}). ProDCARL candidates fall within a dense region of natural AMP embeddings, which indicates that fine-tuning and RL alignment retain global AMP-like sequence semantics. The tighter clustering is expected under the fixed-length constraint ($L=30$) and reward shaping that concentrates sampling toward a narrower subset of AMP-like physicochemical profiles (e.g., cationic and hydrophobic patterns). The observed overlap supports semantic consistency, while the reduced spread indicates that ProDCARL does not capture the full diversity of the natural AMP landscape.

\subsection{Optimization Dynamics and Ablations (Reward vs Diversity)}
\begin{figure} 
    \centering
    \includegraphics[width=\columnwidth]{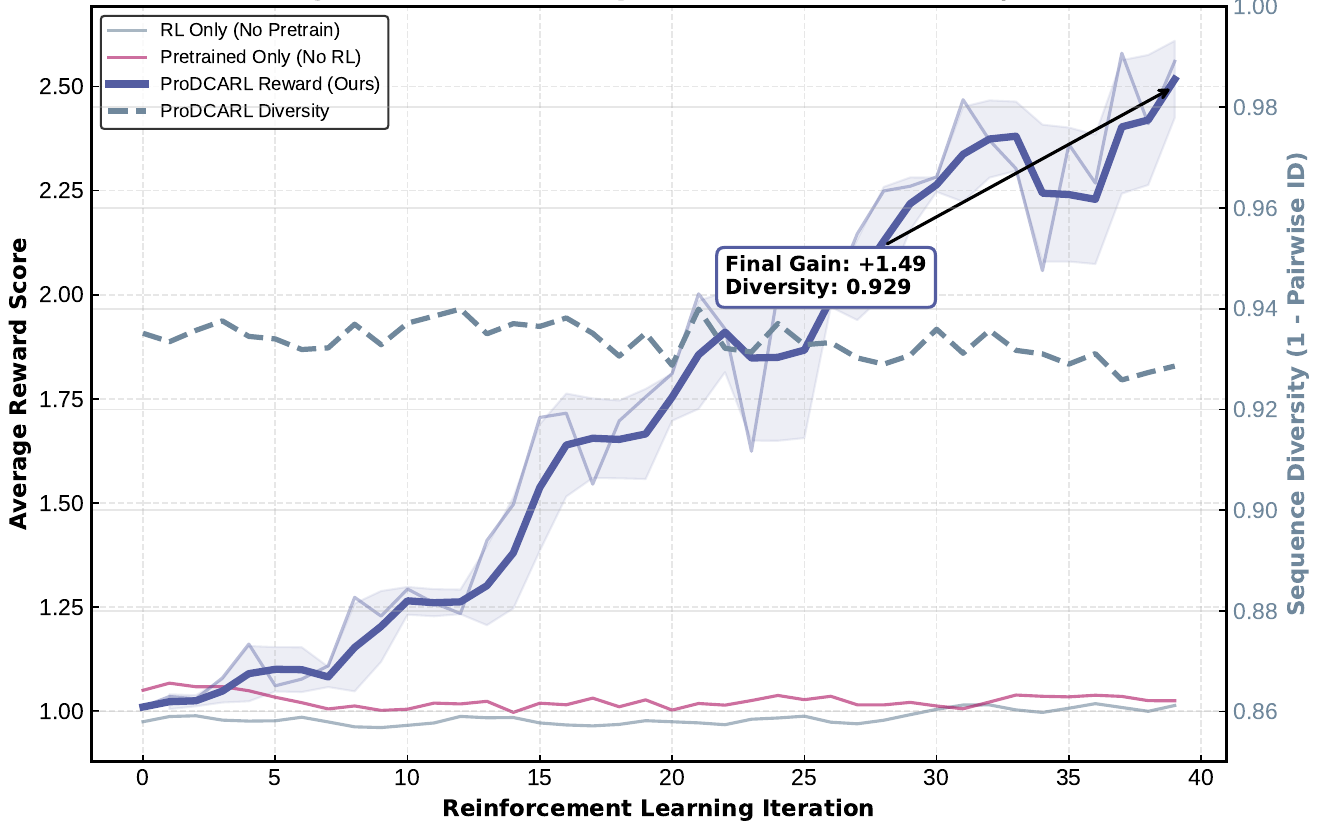}
    \vspace{-3mm}
    \caption{\textbf{Reward and diversity during reinforcement learning.} ``EvoDiff+RL'' denotes RL updates from the base EvoDiff model without AMP fine-tuning. ``EvoDiff+FT'' denotes the AMP fine-tuned prior without RL updates. ProDCARL initializes RL from the AMP fine-tuned prior and applies top-$k$ updates with entropy regularization and early stopping.}
    \label{fig:ablation}
\end{figure}

We conducted an ablation study to isolate the roles of AMP-domain adaptation (fine-tuning) and RL-based alignment. 
Figure~\ref{fig:ablation} shows that ProDCARL increases the average reward while preserving diversity in early and mid training.
In contrast, RL on the pretrained EvoDiff model, without AMP fine-tuning, produces only modest improvements.
This pattern aligns with sampling far from the AMP manifold, where rewards are sparse and gradient estimates become noisy.
Fine-tuning on AMP sequences creates a domain-aware prior that increases the share of AMP-like samples.
As a result, the reward models provide more informative gradients for subsequent RL updates.
Top-$k$ updates stabilize optimization because they focus learning on high-reward samples during each update step.
Entropy regularization and early stopping limit diversity collapse when the reward model receives excessive weight.
We ran RL for up to 50 iterations and report results at the early-stopped iteration when diversity began to decline.

\begin{table*}[htbp]
\centering
\caption{Comparison of sequence generation quality across different methods. All models achieved a Uniqueness score of 1.0.}
\label{tab:generation_comparison}
\resizebox{.7\textwidth}{!}{%
\begin{tabular}{lcccc} 
\toprule
Model & Avg AMP & Avg Tox & AMP $>$ 0.7 \& Tox $<$ 0.3 (\%) & Diversity \\
\midrule
EvoDiff      & 0.038 & 0.061 & 0.3 & 0.923 \\
EvoDiff+FT   & 0.081 & 0.076 & 2.0 & 0.930 \\
EvoDiff+RL   & 0.037 & 0.073 & 0.1 & 0.926 \\
ProDCARL     & \textbf{0.178} & \textbf{0.142} & \textbf{6.3} & \textbf{0.929} \\
\bottomrule
\end{tabular}%
}
\end{table*}

\subsection{Quantitative Screening Performance and Trade-offs}
Table~\ref{tab:generation_comparison} summarizes screening-relevant metrics across all evaluated methods for sequence generation in this study.
Fine-tuning alone improves Avg AMP and the joint hit rate relative to the EvoDiff base model.
This result suggests that domain adaptation raises the chance of sampling AMP-like sequences.
RL without fine-tuning yields little gain, and reward learning works best with a domain-aware prior.
ProDCARL achieves the highest Avg AMP (0.178) and the highest joint hit rate (6.3\%).
ProDCARL maintains high diversity, with $1-\text{mean pairwise identity}=0.929$, which is close to EvoDiff+FT ($0.930$).

Avg Tox increases under ProDCARL when computed across all generated samples in this evaluation. This behaviour is consistent with the known coupling between cationic and hydrophobic motifs that increase membrane activity and motifs that raise toxicity estimates.
Moreover, this pattern follows from strong pressure toward predicted activity, which can steer search toward cationic and hydrophobic sequences with higher toxicity scores.
In practice, ProDCARL serves as a candidate generator, and we apply joint-threshold filtering (pAMP $>$ 0.7 and pTox $<$ 0.3) plus deduplication.
The joint hit rate after these constraints is therefore the main screening metric for practical use.


\subsection{Limitations and Future Work}

First, the reported screening metrics rely on the same AMP and toxicity predictors used as reward models, so the results quantify reward-aligned screening performance rather than independent predictor agreement.

Second, ProDCARL optimizes surrogate reward models, namely AMP and toxicity predictors, rather than experimental assays.
This design can invite reward hacking when the generator exploits predictor preferences without real activity or safety.
Top-k updates, entropy regularization, deduplication, and early stopping reduced collapse in our experiments, but stronger safeguards remain necessary.
Future work should test reward ensembles, uncertainty-aware penalties, and constrained multi-objective formulations to improve robustness.

Third, RL alignment quality depends on dataset coverage, annotation completeness, and the labeling assumptions used during training.
AMP negatives from SwissProt may include unannotated AMPs, and the smaller toxicity set can limit generalization.
Different split strategies and redundancy filters across tasks may also create distribution shift between predictors and generated candidates.
External validation on held-out sources and additional predictors would better test generalization and reduce single-model bias.

Fourth, we restrict generation to a fixed length of $L=30$, which simplifies optimization but narrows coverage of natural AMP lengths.
This constraint may contribute to tighter clustering in representation space and to reduced sequence diversity.
Variable-length generation with conditioning on target length ranges would broaden applicability across peptide families.

Fifth, the reward accounts for predicted activity and toxicity, but it omits key developability factors such as stability and solubility.
It also ignores immunogenicity, protease resistance, and manufacturability, which matter for translation to practice.
Additional objectives or explicit constraints for these properties would support practical candidates and reduce overly hydrophobic or cationic outputs.

Finally, all the results of our study are in silico, so the study provides computational triage rather than biological validation.
ProDCARL prioritizes a small set of candidates for wet-lab testing, where assays determine potency, selectivity, and safety.

\section{Conclusion}
ProDCARL shows that reinforcement learning can steer a diffusion-based peptide generator toward candidates that satisfy joint \textit{in silico} constraints for antimicrobial activity and toxicity and preserve diversity.
To reduce reliance on brute-force wet-lab screening, this approach can support more sustainable and potentially more equitable early-stage antimicrobial discovery pipelines.
ProDCARL serves as a candidate generator that prioritizes sequences for downstream experimental validation.
An important next step is stronger protection against reward hacking through ensemble rewards and constrained multi-objective optimization.


\bibliographystyle{IEEEtran}
\bibliography{Ref}

\end{document}